\documentclass[conference,compsoc]{IEEEtran}
\ifCLASSOPTIONcompsoc
  \usepackage[nocompress]{cite}
\else
  \usepackage{cite}
\fi
\usepackage{amsmath,amssymb,amsfonts}
\usepackage{amsthm}
\usepackage{algorithmic}
\usepackage{graphicx}
\usepackage{textcomp}
\usepackage{xcolor}

\ifCLASSINFOpdf
\else
\fi
\begin{document}
%
\title{Joint wireless and computing resource management with optimal \\ slice selection in in-network-edge metaverse system}

\author{
    \IEEEauthorblockN{
        Sulaiman Muhammad Rashid\IEEEauthorrefmark{1},
        Ibrahim Aliyu\IEEEauthorrefmark{1},
        Abubakar Isah\IEEEauthorrefmark{1},
        Jihoon Lee\IEEEauthorrefmark{1}, \\
        Sangwon Oh\IEEEauthorrefmark{1}, 
        Minsoo Hahn\IEEEauthorrefmark{2} and  
        Jinsul Kim\IEEEauthorrefmark{1}
    }
    \IEEEauthorblockA{\IEEEauthorrefmark{1}Department of Intelligent Electronics and Computer Engineering, Chonnam National University, Gwangju, S/Korea}
    \IEEEauthorblockA{\IEEEauthorrefmark{2}Department of Computational and Data Science, Astana IT University, Astana, Kazakhstan}
    \IEEEauthorblockA{Email: jsworld@jnu.ac.kr}
}


%


\maketitle

\begin{abstract}
This paper presents an approach to joint wireless and computing resource management in slice-enabled metaverse networks, addressing the challenges of inter-slice and intra-slice resource allocation in the presence of in-network computing. We formulate the problem as a mixed-integer nonlinear programming (MINLP) problem and derive an optimal solution using standard optimization techniques. Through extensive simulations, we demonstrate that our proposed method significantly improves system performance by effectively balancing the allocation of radio and computing resources across multiple slices. Our approach outperforms existing benchmarks, particularly in scenarios with high user demand and varying computational tasks.
\end{abstract}


%
\IEEEpeerreviewmaketitle

\section{Introduction}
As technology advances, the new wave of networks is blending virtual and augmented realities with our real world. This creates a whole new space called the metaverse, where the digital and physical worlds come together \cite{9720526}. Creating and maintaining the Metaverse requires enormous resources, including computing resources for intensive data processing to support the Extended Reality, storage resources, and communication resources for maintaining ultra-high-speed and low-latency connections \cite{10158923}.

Given these demands for resources, there is a need for a solution that efficiently manages and allocates these diverse types of resources to power the applications and functionalities within the Metaverse. Network slicing, where multiple end-to-end networks are created on shared physical infrastructure, solves this problem. Different network slices can be used for specific applications or services \cite{7899415}. They can also scale up and down according to the service requirements, and network resources to the metaverse can be allocated according to the demand to facilitate the recommended QoE to the user while optimizing the resource usage in the network.

The metaverse consists of a large-scale number of virtual worlds that require interoperability with each other. These virtual worlds must be rendered in real-time and synchronized with the physical world. The MEC provides a distributed infrastructure located near the user, however it fails to meet concurrent user demands, which results in high delay and severely affects the full realization of the metaverse\cite{aliyu2023dynamic}. According to Internet Research Task Force (IRTF) \cite{kunze2022use}, the Computing in the network (COIN) research group, 6G edge nodes acting as task executors are not only purpose-built servers. They encompass any edge node augmented with computing resources. This offers a promising solution, utilizing untapped network resources for task execution, thereby diminishing latency and fulfilling quality of experience (QoE) requirements. However, augmenting computing resources or enabling COIN increases power consumption \cite{rashid2024graph}. Effectively allocating COIN resources in real-time to adapt to dynamic user demands while ensuring overall system availability still presents a critical challenge.

Despite the advancement in resource management problems, challenges persist. Network slicing and in-network computing have been extensively investigated in the literature to address such challenges. Previous approaches assume each application is dynamically assigned to a specific slice with a static resource pool based on workload and SLA requirements \cite{redana20195g}. However, dynamic assignment leads to mixed workloads in slices and consequently demands flexibility in managing these resources. Jo\v{s}ilo et al.\ \cite{jovsilo2022joint} tackled the resource management and dynamic assignment of slices, by solving complex algorithms in an edge computing system, but with the presence of in-network computing, where multiple nodes act as potential task executors, efficiently managing these resources still remains a challenge.

Unlike the above works, the authors in \cite{sasan2024joint} used a Water-Filling-based heuristic algorithm to address joint network slicing and in-network computing resource allocation. However, their focus was solely on managing resources between slices (inter-slice) without considering the resource management issues within the slices (intra-slice).

In this paper, we address the joint network slicing, inter-slice radio, intra-slice radio and in-network resource management problem and make the following key contributions: 

\begin{itemize}
    \item We formulated the problem as a mixed-integer non-linear programming problem (MINLP) and achieved the optimal solution through a standard optimization solver. 
    \item We performed an extensive evaluation under different load and task settings and compared it against some benchmark solutions.  
\end{itemize}

\section{System Model}

As a reference scenario, we considered a slice-enabled metaverse network architecture characterized by distinct application slices \( \mathcal{N} = \{1, 2, \ldots, N\} \), which have combinations of computing resources optimized for executing metaverse intensive tasks. We denote by \( \mathcal{I} = \{1, 2, \ldots, I\} \) as a set of Wireless Devices (WD) that generate computationally intensive tasks, each with varying computational demands. Task $i$ generated by $I$ characterized with input size $S_i$ can either be computed locally or assigned to a slice $n$ through set \( \mathcal{A} = \{1, 2, \ldots, A\} \) of access points (AP).

\begin{figure}[!ht]
  \centering
  \includegraphics[width= 1.1\linewidth]{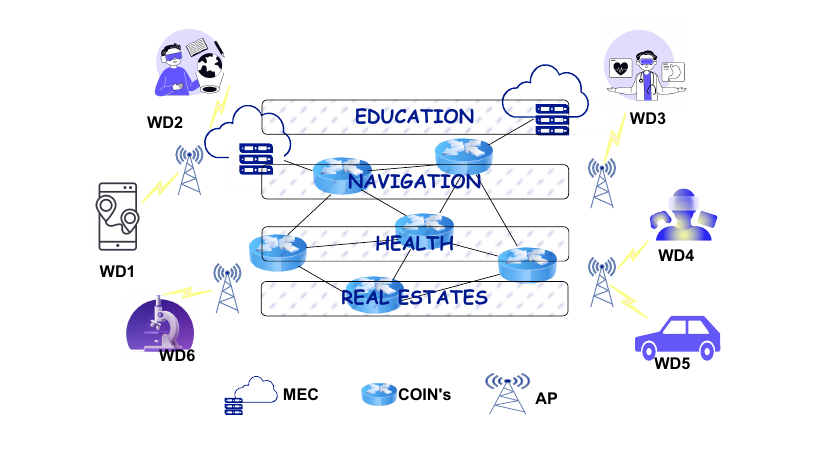}
  \caption{Slice-enabled Reference Scenario}
  \label{fig:architecture}
\end{figure}

In the case of assigning the task to a slice, it goes through exactly one AP and can either be computed within the network on a set of COIN \( \mathcal{C} = \{1, 2, \ldots, C\} \), or can be offloaded to a MEC of set \( \mathcal{M} = \{1, 2, \ldots, M\} \). For simplicity, we denote $\overline{N}$ to be the set of all edge nodes (ENs), i.e. $\mathcal{C} \cup \mathcal{M}$. The APs and ENs form the set $\mathcal{E} \triangleq \overline{N} \cup \mathcal{A}$ of edge resources. For all the tasks generated, there is an expected number of instructions required to perform the computation. Since the WD, COIN and MEC have different instruction sets of architectures, the required number of instructions may also vary. Therefore, for a task generated by WD $i$, we denote by $\mathcal{L}_i$, $\mathcal{L}_{i,n}$ as the expected number of instructions locally and in slice $n$ respectively which we considered them to be estimated by methods described in [ref]. 

We define set of decisions for task $i$ as \( \Delta_i = \{ i \} \cup \{ (a, j, n) \,|\, a \in \mathcal{A}, j \in \overline{N}, n \in \mathcal{N} \} \)  and we will use $d_i$ \(\in\) $\Delta_i$ to indicate the decision for WD $i$'s task i.e $\delta_i = i$ indicates that the task should be performed locally, and $d_i = (a, j, n)$ indicates that task should be offloaded through AP $a$ to node $j$ in slice $n$. Hence, we define a decision vector \( \boldsymbol{\delta} = (\delta_i)_{i \in \mathbb{I}} \) as the collection of the decisions of all WD's, and we define the set \( \Delta = \times_{i \in \mathbb{I}} \Delta_i \), i.e., the set of all possible decision vectors

\subsection{Communication Resources}

In the network, the communication resources are managed at the network and slice levels. At the network level, the radio resources of each access point (AP) are shared across the slices based on an interslice radio resource allocation policy, denoted as \( \Pi_{\omega}: \Delta \rightarrow \mathbb{R}_{[0,1]}^{|\mathcal{A}|\times|\mathcal{N}|} \). This policy determines the inter-slice radio resource provisioning coefficients \( \omega_{n}^{a} \), where \( \forall (a, n) \in \mathcal{A} \times \mathcal{N} \), \( \omega_{n}^{a} \leq 1 \).

At the slice level, the radio resources assigned to each slice are shared among the WD's according to an intraslice radio resource allocation policy \( \Pi^{n}_{\phi_a}: \Delta \rightarrow \mathbb{R}^{|\mathcal{A}|\times|\mathcal{I}|}_{[0,1]} \). This policy determines the intra-slice radio resource provisioning coefficients \( \phi_{i,a}^{n} \in [0,1] \), where \( \forall a \in \mathcal{A} \).

The achievable Physical rate of a WD \( i \) at AP \( a \), denoted as \( R_{i,a} \), captures the expected channel conditions, which can be estimated through historical measurements. Given \( R_{i,a} \) and the provisioning coefficients \( \omega_{a}^{n} \) and \( \phi_{i,a}^{n} \), the uplink rate of WD \( i \) at AP \( a \) in slice \( n \) is expressed as:

\begin{equation}
\mathcal{U}_{i,a}^{n}(\boldsymbol{\delta}, \Pi_b, \Pi_{\phi_a}^{n}) = \omega_{a}^{n} \phi_{i,a}^{n} R_{i,a}
\end{equation}

The uplink rate, along with the input data size \( S_i \), determines the transmission time of WD \( i \) in slice \( n \) at AP \( a \):

\begin{equation}
T^{tx,n}_{i,a}(\boldsymbol{\delta}, \Pi_b, \Pi_{\phi_a}^{n}) = \frac{S_i}{\mathcal{U}_{i,a}^{n}(\boldsymbol{\delta}, \Pi_b, \Pi_{\phi_a}^{n})}
\end{equation}

\subsection{Computing Resources}

In our system model, we distinguish between two main categories of computing resources: edge resources and local resources. Within this framework, each slice \( n \) is equipped with a specific combination of computing resources $\overline{N}$. We denote by \( F_{j}^{n} \) the computing capability of node $j$ in slice $n$.   The allocation of these computing resources among the WD's is controlled by intra-slice computing resource allocation policy \( \Pi^{n}_{\phi_{j}}: \Delta \rightarrow \mathbb{R}^{|\mathcal{\overline{N}}|\times|I|}_{[0,1]}\). This policy determines the provisioning coefficients \( \phi_{i,j}^{n} \in [0,1]\), such that $ \quad \forall (j,n) \in \mathcal{N} \times \overline{N}$ ensuring that each WD receives a portion of the available computing power within the slice. Specifically, the computing capability $F_{j}^{n}$ allocated to WD \( i \) in slice \( n \) is expressed in the following equation: 
\begin{equation}
\label{eqn3}
F_{i,j}^{n}(\boldsymbol{\delta}, \Pi^{n}_{\phi_j}) = \phi_{i,j}^{n} F_{j}^{n} \qquad  j \in \overline{N}
\end{equation}

We assumed, like in the literature \cite{fan2018application} and justified by empirical studies \cite{zhao2017tasks}, that the arrival rate of requests of task generated by WD $i$ at node $j$ of slice $n$ follows a Poisson distribution with parameter $\alpha_i$. The expected number of instructions \( \mathcal{L}_{i,n} \) required to execute a task generated by WD $i$ in slice $n$ is exponentially distributed. Therefore, all nodes can form an M/M/1 queuing model \cite{lia2022} to process its corresponding computing tasks.

Thus, exploiting the queuing theory, the average computation time for a generic task at node $j$ of slice $n$ can be derived as:

\begin{equation}
T^{ex}_{i,j}(\boldsymbol{\delta}, \Pi^{n}_{\phi_{j}}) = \frac{1}{\frac{F_{i,j}^n}{\mathcal{L}_{i,n}} - \sum_{i \in \mathcal{I}} \delta_{ij}\alpha_i}, \qquad  j \in \overline{N}
\end{equation}

To keep the queue stable, the average arrival rate, $\alpha_i$, should be smaller than the average service rate, as described by the following equation:

\begin{equation}
\frac{F_{i,j}^n}{\mathcal{L}_{i,n}} - \sum_{i \in \mathcal{I}} \delta_{ij}\alpha_i > 0, \qquad  j \in \overline{N} 
\end{equation}

In addition to the cloud resources, each WD possesses local computing capabilities denoted by \( F_{i}^{l} \), which may vary over different devices. The local execution latency \( T^{ex}_{i} \) of WD \( i \) is simply expressed as 
\begin{equation}
T^{ex}_{i} = \frac{\mathcal{L}_i}{F^{l}_{i}}
\end{equation}

\subsection{Cost Model}
For $e \in \mathcal{E} \cap A$, we denote that $e$ is a communication resource and $\mathcal{T}_{i,e}^{n}$ is the minimum transmission time that WD $i$ will achieve if it is the only WD offloading its computation through AP $e$ in slice $n$. Similarly, for $e \in \mathcal{E} \cap C$, we have that $e$ is a computing resource and $\mathcal{T}_{i,e}^{n}$ is the minimum execution time that WD $i$ will achieve if it is the only WD offloading its computation to Node $j$ in slice $n$

The cost of WD \( i \) is then determined by the task completion time, considering both local computation and offloading to edge:

\begin{multline}
C_i(\boldsymbol{\delta}, \Pi_b, \Pi_{\phi_a}, \Pi_{\phi_j}) = T^{\text{ex}}_{i} {I}(\delta_i, i) \\
+ \sum_{n \in N} \sum_{j \in \overline{N}} \sum_{a \in A} 
\left( \frac{\mathcal{T}^{n}_{i,a}}{\omega_{a}^{n} \phi_{i,a}^{n}} 
+ \frac{\mathcal{T}^{n}_{i,j}}{\phi_{i,j}^{n}} \right){I}(\delta_i, (a, j, n))
\end{multline}

Here, \( (\Pi_{\phi_a}, \Pi_{\phi_j}) = ((\Pi_{\phi_a}^{n}, \Pi_{\phi_j}^{n},)) n \in N \)  represents the collection of slice policies.

The cost of each slice \( n \) is calculated as the sum of transmission and execution times for all WD's offloading their tasks in slice \( n \):

\begin{equation}
C_{(n)}(\boldsymbol{\delta}, \Pi_b, \Pi^n_{\phi_a}, \Pi^n_{\phi_j}) = \sum_{e \in \mathcal{E}} \sum_{i \in \delta_{i,j}} \frac{\mathcal{T}_{i,e}^n}{\omega_e^n \phi_{i,e}^{n}} 
\end{equation}

Here, \( \omega_e^n = 1 \) if \( e \) is a computing resource.

Finally, the system cost is expressed as the sum of individual WD costs and slice costs:

\begin{equation}
\begin{split}
\mathcal{C}(\boldsymbol{\delta}, \Pi_b, \Pi_{\phi_a}, \Pi_{\phi_j}) = & \sum_{n \in N} \mathcal{C}_{(n)}(\boldsymbol{\delta}, \Pi_b, \Pi^n_{\phi_a}, \Pi^n_{\phi_j}) \\
& + \sum_{i \in \delta_i} \mathcal{C}_i^l
\end{split}
\end{equation}

This system cost formulation accounts for both local and offloaded computation across slices.

\section{Optimization Problem Formulation}

We aim to minimize the system cost by finding the optimal decision from vector $\boldsymbol{\delta}$ of offloading decisions. From the above analytical results, the problem can be expressed mathematically as a mixed-integer non-linear programming (MINLP) problem and is formulated as follows:

\begin{equation}
\label{Optimal equation}
\begin{aligned}
& \underset{\boldsymbol{\delta}, \Pi_b, \Pi_{\phi_a} , \Pi_{\phi_j}}{\text{min}} \quad \mathcal{C}(\boldsymbol{\delta}, \Pi_b, \Pi_{\phi_a} , \Pi_{\phi_j} ) 
\end{aligned}
\end{equation}

s.t.

\begin{align}
\text{} & \quad \sum_{\delta \in \mathcal{D}_i} {I}(\delta_i, \delta)=1, \quad i \in \mathcal{I} \tag{10a} \label{eq:C1} \\
\text{} & \quad T^{ex}_{i,j}(\boldsymbol{\delta}, \Pi^{n}_{\phi_{j}}) \leq T^{ex}_{i}, \quad i \in \mathcal{I} \tag{10b} \label{eq:C2} \\
\text{} & \quad \frac{F_{i,j}^n}{\mathcal{L}_{i,n}} - \sum_{i \in \mathcal{I}} \delta_{ij}\alpha_i > 0, \quad  j \in \overline{N}  \tag{10c} \label{eq:C3} \\
\text{} & \quad \sum_{n \in N} \omega_{a}^{n} \leq 1, \quad a \in \mathcal{A} \tag{10d} \label{eq:C4} \\
\text{} & \quad \sum_{i \in  O_{e,n}(\boldsymbol{\delta})}\phi_{i,e}^{n} \leq 1, \quad n \in \mathcal{N}, e \in \mathcal{E} \tag{10e}
\label{eq:C5} 
\end{align}

Constraint \ref{eq:C1} ensures that each WD performs computation either locally or offloads its task to exactly one logical resource in the edge $(a, j, n) \,|\, a \in \mathcal{A}, j \in \overline{N}, n \in \mathcal{N}$. Constraint \ref{eq:C2} indicates that task completion time when offloading is not greater than when computing locally. Constraint \ref{eq:C3} forces the average service rate of edge nodes to be greater than the average task arrival rate in the case of offloading. Constraints \ref{eq:C4} \& \ref{eq:C5} enforces limitations on the amount of radio resources that can be provided to an AP in each slice and the amount of computing resources of an edge node that can be provided to each WD in each slice.

\section{Numerical Results}

We conducted extensive simulations to evaluate the performance of our proposed solution, considering individual slices, devices, and channel access points. The simulation scenario involved a single MEC and 8 randomly placed COINs, with 3 AP's and a number of WD sending task requests simultaneously. The euclidean distance between WD$_i$ and node $j$ are set to 2 and 4 for COIN and MEC respectively according to the path loss model in \cite{saunders2007antennas}. The details of the simulation parameters used are explained in Table 1. 


We compare our inter-slice allocation policy $\Pi_b$ to another policy $\Pi_{b}^{e}$, where the bandwidth of each AP $a$ is shared equally among the slices. We obtained our results through ipopt \cite{wachter2002ipopt} and gurobi \cite{gurobi} standard optimization solvers and pyomo \cite{bynum2021pyomo} framework from the average of 1000 simulations with 90\% confidence intervals.

We consider the system performance by varying the number of slices. To do so, we evaluate the system cost reached by optimal intra-slice allocation policies ($\Pi_{\phi_a}, \Pi_{\phi_j}$) and determine the performance gain $G$ as the ratio between the system cost attained under policy $\Pi_{b}^{e}$ to system cost reached under policy $\Pi_b$.

\begin{figure}[!ht]
  \centering
  \includegraphics[width= 1.0\linewidth]{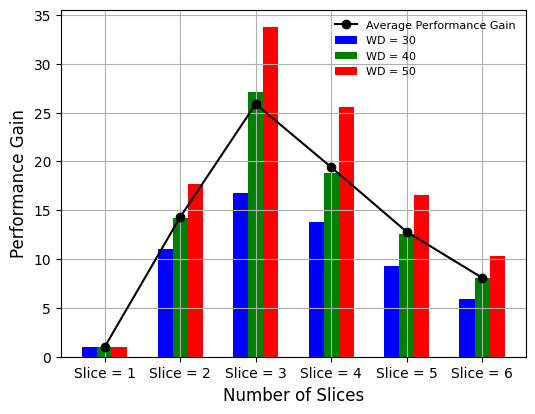}
  \caption{Performance gain vs Number of slices}
  \label{performance gain 1}
\end{figure}

\begin{table}[h!]
\centering
\begin{tabular}{|l|l|}
\hline
\textbf{Parameter} & \textbf{Settings} \\ \hline
\textbf{WD} & \begin{tabular}[c]{@{}l@{}}
- Processing power: [2, 45.4] GIPS \\
- Transmission power: [10e$^{-6}$, 0.1] W \\
\end{tabular} \\ \hline
\textbf{COIN} & - Processing power: [72, 768] GIPS \\ \hline
\textbf{MEC} & - Processing power: 1285 GIPS \\ \hline
\textbf{Task Size} & [1.7, 10] MB \\ \hline
\textbf{AP Bandwidth} & (18, 27) MHz \\ \hline
\end{tabular}
\vspace{0.5em} 
\caption{Simulation Parameters and Settings}
\label{tab:simulation_params}
\end{table}
In Figure \ref{performance gain 1}, we observe that for every number of WD's, $G$ remains 1 when Slice = 1, indicating that the two policies are equivalent when there is no slicing. However, when Slice $>$ 1, $G > 1$ and continues to increase as the number of WD increases, this is because the policy $\Pi_{b}^{e}$ does not account that the slices might have different amount of both radio and computing resources. At Slice = 3, we achieve the highest performance gain across all numbers of WD. This suggests that the policy $\Pi_{b}$ optimally balances the allocation of radio and computing resources at this point. However, when Slice $>$ 3, the performance gain begins to decline, which shows that beyond 3 slices, the overhead of managing additional slices and the resulting resource allocation outweigh the benefits of further slicing for the given resource availability in the initial setup.

\begin{figure}[!ht]
  \centering
  \includegraphics[width= 1.0\linewidth]{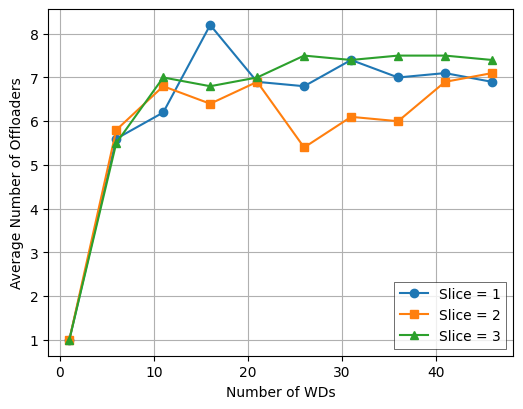}
  \caption{Number of offloaders per slice vs number of WD's}
  \label{performance gain 2}
\end{figure}

 Figure \ref{performance gain 2} shows how offloading behaviour changes as the number of WD's increases under different slicing scenarios. As the number of WD's increases beyond 10, the average number of offloaders stabilizes across all configurations. This implies that the system reaches a point where the optimal policies $(\Pi_b, \Pi^n_{\phi_a}, \Pi^n_{\phi_j})$ effectively balance the available computing resources in the slices with the WDs' preferences for different types of computing resources. 

\section{Conclusion}
This paper addresses the critical challenge of joint wireless and computing resource management in a slice-enabled metaverse network. By formulating the problem as a mixed-integer nonlinear programming (MINLP) problem, we provided an optimal solution that effectively manages both inter-slice and intra-slice resources. The proposed method demonstrates significant improvements in system performance, particularly in scenarios with multiple slices, highlighting the importance of tailored resource allocation strategies.

In future work, we propose integrating artificial intelligence (AI) to enhance the efficiency and adaptability of resource management in metaverse networks.


\ifCLASSOPTIONcompsoc
  \section*{Acknowledgment}
\else
  \section*{Acknowledgment}
\fi

This work was partly supported by Innovative Human Resource Development for Local Intellectualization program through the Institute of Information \& Communications Technology Planning \& Evaluation(IITP) grant funded by the Korea government(MSIT) (IITP-2024-RS-2022-00156287, 50); and This work was partly supported by the Institute of Information \& Communications Technology Planning \& Evaluation (IITP) grant funded by the Korean government (MSIT) No.RS-2021-II212068 Artificial Intelligence Innovation Hub.



%
\bibliographystyle{ieeetr}
\bibliography{bibliography}

\end{document}